\documentclass[12pt]{article}
\usepackage{graphicx}

\newsavebox{\sboxpubnumber}
\newsavebox{\sboxpubdate}
\newcommand{\pubdate}[1]{\begin{lrbox}{\sboxpubdate}{#1}\end{lrbox}}
\newcommand{\pubnumber}[1]{\begin{lrbox}{\sboxpubnumber}{\begin{tabular}{l} #1 \\
				 \usebox{\sboxpubdate}
				 \end{tabular}}
                           \end{lrbox}
                           \pubblock}
\newcommand{\Title}[1]{\begin{center} {\Large #1 } \end{center}}
\newcommand{\Author}[1]{\begin{center}{ \sc #1} \end{center}}
\newcommand{\Address}[1]{\begin{center}{ \it #1} \end{center}}
\newcommand{\andauth}{\begin{center}{and} \end{center}}
\newcommand{\pubblock}{\rightline{
			\usebox{\sboxpubnumber}}}
\newenvironment{Abstract}{\begin{quotation}  }{\end{quotation}}
\newenvironment{Presented}{\begin{quotation} \begin{center}
             PRESENTED AT\end{center}\bigskip
      \begin{center}\begin{large}}{\end{large}\end{center}
      \end{quotation}}
\newcommand{\Acknowledgements}{\bigskip  \bigskip \begin{center} \begin{large}
             \bf ACKNOWLEDGEMENTS \end{large}\end{center}}
\newcommand{\gsim}{
\raisebox{0.2em}{$>$} \hspace{-0.75em} \raisebox{-0.2em}{$\sim$}
 }
\newcommand{\lsim}{
\raisebox{0.2em}{$<$} \hspace{-0.75em} \raisebox{-0.2em}{$\sim$}
 }

\begin{document}

\begin{titlepage}
\pubdate{\today}                    
 \pubnumber{KOBE-TH-01-08} 

\vfill
\Title{Stability and experimental flux bound of Fermi Ball}
\vfill
\Author{Kenzo Ogure\footnote{This author is a research fellow of the
 Japan Society for the Promotion of Science(No.4834).}}
\Address{Department of Physics, Kobe University, \\
         Rokkoudaicho 1-1, Nada-Ku, Kobe 667-8501, Japan}
\vfill
\andauth
\vfill
\Author{Jiro Arafune}
\Address{ National Institution for Academic Degrees,\\
         Hitotsubashi 2-1-2, Chiyoda-Ku, Tokyo 101-8438, Japan}
\Author{Takufumi Yoshida}
\Address{Department of Physics, Tokyo University,\\
        Hongo 7-3-1, Bunkyo-Ku, Tokyo 113-0033, Japan}
\vfill
\begin{Abstract}
We investigate the stability of an Fermi ball(F-ball) within the
 next-to-leading order approximation in the thin wall expansion.  We
 find out that an F-ball is unstable in case that it is electrically
 neutral.  We then find out that an electrically charged F-ball is
 metastable in some parameter range.  We lastly discuss the allowed
 region of parameters of an F-ball, taking into account the stability of
 an F-ball and results of experiments.
\end{Abstract}
\vfill
\begin{Presented}
    COSMO-01 \\
    Rovaniemi, Finland, \\
    August 29 -- September 4, 2001
\end{Presented}
\vfill
\end{titlepage}
\def\thefootnote{\fnsymbol{footnote}}
\setcounter{footnote}{0}

\section{Introduction}

Fermi ball(F-ball), which is a kind of nontopological solitons, is first
proposed as a candidate for cold dark matter(CDM) \cite{Mac}.  Similar
object is considered to have possibility to explain the baryon number
asymmetry of the present universe \cite{Mor}.  An F-ball consists of a
closed domain wall and fermions which localize on the wall. Such an
object is thought to be produced after a phase transition, in which two
almost degenerate vacua exist.  At this phase transition, two regions,
which correspond to two vacua, and domain walls are produced.  If these
two vacua are completely degenerate, the domain walls will dominate the
energy density of the universe soon \cite{Vil}.  However, if the energy
density of one vacuum is slightly smaller than the other one, the true
vacuum pushes the domain walls and reduces the false vacuum region.  If
some fermions are captured on the domain walls, the left region may be
stabilized due to the Fermi pressure of the captured fermions.

  We first assume that the region has complete spherical form with
radius, $R$.  In this case, the total energy of this system consists of
three parts, the surface, the volume, and the Fermi energy within the
thin wall approximation:
\begin{eqnarray}
 E_{total}
=
E_s + E_v + E_F
=
4\pi R^2 \Sigma + \frac{4\pi R^3 \epsilon}{3} + \frac{2 N^\frac{3}{2}}{3 R}.
\end{eqnarray}
Here, $\Sigma$, $\epsilon$, and $N$ are the surface tension, the energy
density difference between the two vacua, and the number of fermions on
the wall, respectively.  Since the surface energy and the volume energy
are the increasing functions of the radius and the Fermi energy is the
decreasing one, this object is stabilized at a certain radius.  We deal
the volume energy as a perturbation term hereafter, since it is much
smaller than the other energies in most cases.  Minimizing this total
energy with respect to $R$, we find out that it is proportional to the
number of the fermions on the wall with the following critical radius,
$R_c$:
\begin{eqnarray}
  E_{total}
=
\frac{1}{3}\kappa N ,
\ \ \ 
(R_c=\frac{\sqrt{N}}{\kappa},\ \kappa \equiv (12 \pi \Sigma)^{\frac{1}{3}}).
\end{eqnarray}
This object is called as Fermi ball.

If this large F-ball were stable\footnote{ We call an F-ball large if
its radius is much larger than the thickness of the wall.  }, it were
interesting as a candidate for CDM.  It, however, is unstable against
deformation from spherical shape and fragments into small pieces
\cite{Mac}.  This can be understood as follows.  Since the energy of an
F-ball is proportional to the number of the fermions on the wall as
above, it has same energy even if it is divided into some smaller
F-balls.  However, there exists the small volume energy, which we
neglected.  Taking into account this contribution, we can easily find
out that the fragmented state has smaller energy.  An F-ball therefore
continues to be small until the thickness of the wall becomes comparable
to its radius and the thin wall approximation breaks down.  Such an
F-ball is too small and there must be large number of them in order that
they have sizable contribution to CDM.  An F-ball was, therefore,
thought to consist of a domain wall and a fermion which interact with
ordinary matter only very weakly \cite{Mac}.

An electrically charged F-ball, which is stable against the
fragmentation owing to the long range electric force, is then proposed
\cite{Mor}.  This F-ball can be large and heavy enough to have sizable
contribution to CDM even if they are very dilute.

We, however, have three questions about the above treatments of an
F-ball:
\begin{itemize}
 \item Is an electrically neutral F-ball unstable even if the curvature
       effect is taking into account?
 \item Is an electrically charged F-ball really stable?
 \item Is an electrically charged F-ball stable even in the hot early
       universe where the F-ball is consider to be produced?
\end{itemize}

  First, when we consider an electrically neutral F-ball, we used the
thin wall approximation and neglected the contribution from the
curvature of the wall.  Though this contribution is really much smaller
than the total energy of an F-ball, it can be very important.
Neglecting the curvature effect, we concluded that the energy of an
F-ball is same as that of its fragmented state in the absence of the
volume energy.  However, if we take into account the contribution from
the curvature, two states will have different energies.  This curvature
contribution, therefore, would determine which state is stable.

Second, The total energy of an electrically charged F-ball approximately
consists of the surface energy and the coulomb energy for large $N$
because the coulomb energy is proportional to $N^2$:
\begin{eqnarray}
 E_{total}
\simeq
E_s + E_c
=
4\pi R^2 \Sigma
+
\frac{\alpha N^2}{2 R}.
\end{eqnarray}
Minimizing the total energy with respect to the radius, we obtain the
total energy as follows:
\begin{eqnarray}
  E_{total}
=
\frac{1}{3} \kappa (\frac{3\alpha}{4})^{\frac{2}{3}} N^{\frac{4}{3}},
\ \ \ 
(R_c=\kappa (\frac{3 \alpha}{4})^{\frac{1}{3}} N^{\frac{2}{3}}).
\end{eqnarray}
Since the total energy is proportional to $N^{\frac{4}{3}}$, a
fragmented state has apparently smaller energy than an F-ball.  We
therefore can not conclude that the electric force stabilize the F-ball
from the fragmentation only by comparing the energies of the two states.
We then should investigate that is there any energy barrier to deform an
F-ball to a fragmented state.  If there is enough barrier due to the
electric long range force, an F-ball can be metastable.

Third, in thermal bath, the electric long range force becomes short
range one with typical Debye screening length.  The energy barrier,
which arises from the electric long range force, would become week or
disappear due to this Debye screening.  

To answer these three questions is the main aim of the present paper.
Details of the following analysis are shown in Ref.\cite{Ara}.

\section{Stability of a neutral F-ball}

We first consider the curvature effect to the stability of an F-ball.
Though we only consider a simple model described by the following
Lagrangian density, most results will be able to be applied to more
complicated models \cite{Mac}:
\begin{eqnarray}
 {\cal L}=\frac{1}{2}(\partial_{\mu}\phi)^2+\bar{\psi}_F(i\gamma^\mu\partial_{\mu}-G\phi)\psi_F-U(\phi)~,
\end{eqnarray}
where $\phi$ and $\psi_F$ are a scalar and a fermion field,
respectively, and $G$ is a Yukawa coupling constant. Here, $U(\phi)$ is
an approximate double-well potential \footnote{Note that the qualitative
discussions in the following do not depend on the explicit form of
$U(\phi)$.},
\begin{equation}
U(\phi)=\frac{\lambda}{8}\left(\phi^2-v^2\right)^2+U_{\epsilon}(\phi)~,
\label{neu_phi4.eqn}
\end{equation}
taking the second term, which breaks $Z_2$ symmetry under $\phi
\leftrightarrow -\phi$, much smaller than the first one. This model has
a kink solution\footnote{ Though we neglect the effect of $U_{\epsilon}$
to show the following explicit solution, the presence of the solution is
not affected by $U_{\epsilon}$.  }, which interpolates the two vacua,
$\phi = v,-v$:
\begin{equation}
\phi (z)=v\tanh{\frac{z-z_0}{\delta_N}}~,
\ \ \ \ 
(\delta_N \equiv 2/( \sqrt{\lambda}v)).
\label{neu_phiz.eqn}
\end{equation}
We assume the width of the wall, $ \delta_N $ is much smaller than the
F-ball size. We call the plane where $\phi$ vanishes "surface" of the
domain wall in the present paper.  This domain wall has the surface
tension, $\Sigma$:
\begin{eqnarray}
 \Sigma
  =
  \frac{2\sqrt{\lambda}}{3}v^3.
\label{surfacetension}
\end{eqnarray}
 Substituting Eq.(\ref{neu_phiz.eqn}) into the Euler-Lagrange
equation of $\psi_F$, we find that $\psi_F$ has a zero-mode solution,
\begin{equation}
\psi_F(z)=\psi_F(z_0)\exp{\left\{-G\int_{z_0}^{z}\mbox{d}z'~\phi(z')\right\}}~.
\label{neu_psiz.eqn}
\end{equation}
Here, $\psi_F(z_0)$ is a spinor eigenstate of $ i\gamma_3=-1 $.  A
fermion can localize on the domain wall owing to this zero mode.  This
model is therefore a suitable model to investigate the properties of an
F-ball, since it has the domain wall solution and the fermion, which can
localize on the wall.

\begin{figure}[htb]
    \centering
    \includegraphics[height=1.5in]{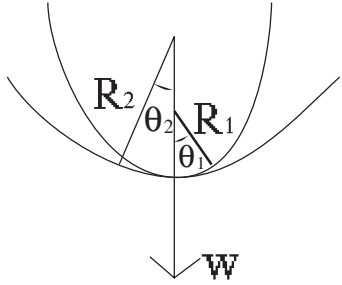}
    \caption{Local coordinates on the wall.}
    \label{localcordinate}
\end{figure}

We evaluate the next-to-leading order approximation of the thin wall
expansion in this model.  We assume that the scalar field expectation
value does not change along the surface and write the surface energy as
follows\footnote{ We only consider cases, $R_1,R_2 > 0$, since it is
sufficient to analyze the stability of an F-ball.  }:
\begin{eqnarray}
 E_s[\phi]
&=&
\int \mbox{d}^3{\bf x}~ \left\{\frac{1}{2}(\nabla\phi({\bf x}))^2
+\frac{\lambda}{8}\left(\phi({\bf x})^2-v^2\right)^2 \right\} \nonumber \\
&\sim &
\int \mbox{d}\theta_1  \mbox{d}\theta_2 
\mbox{d}w~ (R_1 +w)(R_2+w) 
 \left\{ \frac{1}{2}\dot{\phi}^2+\frac{\lambda}{8} 
\left( \phi^2-v^2 \right)^2 \right\}~\\
&=&
\int \mbox{d}S \mbox{d}w~ (1+\frac{w}{R_1})(1+\frac{w}{R_2}) 
 \left\{ \frac{1}{2}\dot{\phi}^2+\frac{\lambda}{8} 
\left( \phi^2-v^2 \right)^2 \right\}~,
\label{neu_etot.eqn}
\end{eqnarray}
where $ \dot{\phi} $ stands for the derivative of $ \phi $ with respect
to $ w $.  The local coordinate, $w$ is perpendicular to the surface and
becomes zero on the surface.  The curvature radiuses, $R_1$ and $R_2$
are principle curvature radiuses on each points and the local
coordinates, $\theta_1$ and $\theta_2$ are the corresponding angular
coordinates (see Fig.\ref{localcordinate}).

Taking variation of $ E_N $ with respect to $ \phi $, we obtain the
equation,
\begin{equation}
\ddot{\phi}+\left( \frac{1}{R_1+w}+\frac{1}{R_2+w} \right)\dot{\phi}=\frac{\lambda}{2}\phi(\phi^2-v^2),
\label{neu_eqM.eqn}
 \end{equation}
with the boundary condition,
\begin{eqnarray}
\phi = \left\{ 
\begin{array}{ll}
\pm v & (w \rightarrow \pm \infty) \\
0 & (w = 0)~~~.
\end{array}
\right.
\label{neu_bc.eqn}
\end{eqnarray}

We now estimate $ E_s $ within the thin-wall approximation, ignoring
the curvature effect.\footnote{This of course leads to the same result as
Macpherson and Campbell derived. Our derivation is however meaningful as
a step toward the estimation of the energy up to the next-to-leading order
contribution in the thin-wall expansion.} The equation (\ref{neu_eqM.eqn})
becomes in the leading order as,
\begin{equation}
\ddot{\phi_0}=\frac{\lambda}{2}\phi_0(\phi_0^2-v^2),
\label{neu_eqM0.eqn}
\end{equation}
with the boundary condition,
\begin{eqnarray}
\phi_0 = \left\{ 
\begin{array}{ll}
\pm v & (w \rightarrow \pm \infty) \\
0 & (w = 0)~~~.
\end{array}
\right.
\label{neu_bc0.eqn}
\end{eqnarray}
The solution to Eq.(\ref{neu_eqM0.eqn}) is the kink,
\begin{equation}
\phi_0(w)=v\tanh{\frac{w}{\delta_N}}~.
\label{neu_phi0.eqn}
\end{equation}

Adding the Fermi energy, we obtain the total energy of the neutral
F-ball within the thin wall approximation, $E_N^0$:
\begin{equation}
E_N^0=\Sigma S+\frac{4\sqrt{\pi}}{3}\int\mbox{d}S~n_F^{3/2}({\bf x}_F)~,
\label{neu_en0.eqn}
\end{equation}
where $ \Sigma $ is the surface tension in Eq.(\ref{surfacetension}).  We
should minimize this total energy, keeping the total fermion number,
\begin{equation}
N=\int\mbox{d}S~n_F({\bf x}_F)~,
\label{neu_nf.eqn}
\end{equation}
constant.  Using the Lagrange's multiplier method, we find that
$n_F({\bf x}_F)$ is constant:
\begin{equation}
n_F({\bf x}_F)=\frac{N}{S}~.
\label{neu_uniform_distr.eqn}
\end{equation}
From this, we finally get the total energy:
\begin{equation}
E_N^0=\Sigma S+\frac{4\sqrt{\pi}N^{3/2}}{3\sqrt{S}}~.
\label{neu_en0.2.eqn}
\end{equation}
Minimizing with respect to $S$, we obtain,
\begin{equation}
E_N^0=(12 \pi \Sigma)^{1/3}N~,
\label{neu_enmin0.eqn}
\end{equation}
with the critical area of the surface,
\begin{equation}
S=\left( \frac{2\sqrt{\pi}}{3\Sigma} \right)^{2/3}N~.
\label{neu_smin0.eqn}
\end{equation}
Since the total energy is proportional to the number of the fermion, we
can not know the stability of the F-ball in this order approximation.

We then calculate the next-to-leading order approximation of the thin
wall expansion.  We here expand $ \phi $ and $ E_N $ with respect to $
\delta_N/R_1 $ and $ \delta_N/R_2 $,
\begin{eqnarray}
\phi&=&\phi_0+\phi_1+\cdots \nonumber\\
E_N&=&E_N^0+E_N^1+E_N^2+\cdots ~.
\label{neu_expand.eqn}
\end{eqnarray}
Since $ E_N^1 $ vanishes owing to the variational principal, we estimate
the energy up to $ E_N^2 $. It is enough to expand $ \phi $ up to $
\phi_1 $ in this case. From Eq.(\ref{neu_eqM.eqn}), $ \phi_1 $ satisfies,
\begin{equation}
\ddot{\phi_1}-\frac{\lambda}{2}\left( 3\phi_0^2-v^2 \right)\dot{\phi_1}=-\left(\frac{1}{R_1}+\frac{1}{R_2}\right)\dot{\phi_0},
\label{neu_eqM1.eqn}
\end{equation}
with the boundary condition,
\begin{eqnarray}
\phi_1 = \left\{ 
\begin{array}{ll}
0 & (w \rightarrow \pm \infty) \\
0 & (w = 0)~~~.
\end{array}
\right.
\label{neu_bc1.eqn}
\end{eqnarray}
The solution to Eq.(\ref{neu_eqM1.eqn}) is,
\begin{equation}
\phi_1(w)=\frac{1}{2\sqrt{\lambda}}\left(\frac{1}{R_1}
+\frac{1}{R_2}\right)f_1(w/\delta_N)~,
\label{neu_phi1.eqn}
\end{equation}
with,
\begin{eqnarray}
f_1(x)&=&\mbox{sech}^2x-\cosh^2{x}+\frac{1}{3}\mbox{sinh}^2x \mbox{tanh}^2x \nonumber \\
&&+\frac{\mbox{sech}^2x}{12} \left| \sinh{4x}+8 \sinh{2x}+12x \right|~.
\label{neu_phi1_2.eqn}
\end{eqnarray}
Substituting $ \phi_0 $ and $ \phi_1 $ into $ E_N $, we obtain,
\begin{equation}
E_N^2=C_N^1 \int \mbox{d}S~\frac{1}{ R_1R_2 } +C_N^2 \int 
\mbox{d}S
\left(\frac{1}{R_1}-\frac{1}{R_2}\right)^2~,
\label{neu_en2.eqn}
\end{equation}
with,
\begin{eqnarray}
C_N^1&=&\frac{1}{\sqrt{\lambda}} \int \mbox{d}w~ 
\left( \sqrt{\lambda}w^2 \dot{\phi_0}^2-\dot{\phi_0}f_1 \right) 
\simeq -\frac{0.25v}{\sqrt{\lambda}} < 0\nonumber \\
C_N^2&=&-\frac{1}{4\sqrt{\lambda}} \int \mbox{d}w~ \dot{\phi_0}f_1 
\simeq -\frac{0.28v}{\sqrt{\lambda}}~ < 0.
\label{neu_cn.eqn}
\end{eqnarray}
We consider the stability against the fragmentation from the
next-to-leading order contribution, $ E_N^2 $.  The first term in
Eq.(\ref{neu_en2.eqn}) is proportional to the integration of Gaussian
curvature on the closed surface, which is known to be $ 4\pi $
(Gauss-Bonnet's Theorem).  We therefore find out that the first term
does not depend on the shape of the F-ball and depend only on the number
of the F-ball.  Since $C_N^1$ is negative, a fragmented state has
smaller energy.  On the other hand, the second term is zero only for
sphere and negative for any other shapes.  A spherical F-ball then
deforms and fragments to small pieces.  We conclude that a neutral
F-ball is unstable even in the absence of the volume energy due to the
curvature effect.

\section{Stability of an electrically charged F-ball}

We consider the stability of an electrically charged F-ball in thermal
bath in the present section.  We here consider the free energy of the
fermion gas on the wall and the electron gas, which surrounds the
F-ball\footnote{ Though we assume that the fermion on the wall has an
electric charge, we can easily extent our analysis to other cases.  }:
\begin{eqnarray}
F_C[V_e]&=& \int \mbox{d}^3{\bf x}~ \left\{ -\frac{1}{2e^2} 
\left( \nabla V_e({\bf x}) \right)^2+{\cal F}_e(V_e({\bf x})) 
\right. \nonumber \\
&& \hspace{2cm} 
\left. \frac{}{} 
-n_F({\bf x}_F)V_e({\bf x})\delta(|{\bf x}-{\bf x}_F|) \right\}~,
\label{scr_fctot.eqn}
\end{eqnarray}
where $ {\cal F}_e $ is the free energy density of the electron
gas. When the temperature, $ T $ is much higher than the electron mass,
$ {\cal F}_e $ can be written as,
\begin{eqnarray}
{\cal F}_e(V_e)
&=&
-\frac{2T}{(2\pi)^3} \int \mbox{d}^3{\bf p}~ 
\left\{\frac{p}{T}+\log{ \left(1+e^{-\frac{p-V_e}{T}} \right)} 
\right. \nonumber \\
&&\hspace{3cm} \left. +\log{ \left(1+e^{-\frac{p+V_e}{T}} \right)} 
\right\} \nonumber \\
&= & -\frac{T^2V_e^2}{6}-\frac{V_e^4}{12\pi^2}+ (V_e\ independent\ terms).
\label{scr_fe.eqn}
\end{eqnarray}

The free energy can be obtained by extremizing Eq.(\ref{scr_fctot.eqn})
with respect to $ V_e({\bf x}) $.  Since $ V_e({\bf x}) $ is expected to
change rapidly near the wall, we also use the thin wall expansion here.
Assuming that $ V_e $ depends only on $ w $, we can express $ F_C $ as,
\begin{eqnarray}
F_C[V_e]&=& \int \mbox{d}S \mbox{d}w~
 (1+\frac{w}{R_1})(1+\frac{w}{R_2}) 
 \left\{ -\frac{1}{2e^2} \dot{V_e}^2-\frac{T^2V_e^2}{6} \right. \nonumber \\
&& \hspace{3cm} \left. -\frac{V_e^4}{12\pi^2}-n_FV_e\delta(w) \right\}~.
\label{scr_fc.eqn}
\end{eqnarray}
Taking variation of Eq.(\ref{scr_fc.eqn}) with respect to $ w $, we
obtain the equation,
\begin{equation}
\frac{1}{e^2} \ddot{V_e}+\frac{1}{e^2} \left( \frac{1}{R_1+w}+\frac{1}{R_2+w} \right)\dot{V_e}=\frac{T^2V_e}{3}+\frac{V_e^3}{3\pi^2}+n_F\delta(w)~,
\label{scr_eqM.eqn}
\end{equation}
with the boundary condition,
\begin{equation}
V_e \rightarrow 0 \hspace{1cm} (w \rightarrow \pm \infty)~.
\label{scr_bc.eqn}
\end{equation}
This corresponds to the well-known Thomas-Fermi equation.

We first calculate $ F_C $ within the thin-wall approximation, ignoring
the curvature effect. From Eq.(\ref{scr_eqM.eqn}), we obtain the equation,
\begin{equation}
\frac{1}{e^2} \ddot{V_e^0}=\frac{T^2V_e^0}{3}
+\frac{(V_e^0)^3}{3\pi^2}+n_F\delta(w)~,
\label{scr_eqM0.eqn}
\end{equation}
with the boundary condition,
\begin{equation}
V_e^0 \rightarrow 0 \hspace{1cm} (w \rightarrow \pm \infty)~.
\label{scr_bc0.eqn}
\end{equation}
The solution of Eq.(\ref{scr_eqM0.eqn}) is
\begin{equation}
V_e^0(w)=\frac{-\sqrt{2}\pi T}{\sinh{ \left( \frac{|w|}{\lambda_T}+c_T \right)}}~,
\label{scr_v0.eqn}
\end{equation}
with
\begin{equation}
\lambda_T=\frac{\sqrt{3}}{eT}~,
\label{scr_lambdat.eqn}
\end{equation}
and with,
\begin{eqnarray}
c_T= \mbox{cosh}^{-1}\frac{1+\sqrt{1+\bar{\sigma}^2}}{\bar{\sigma}}, 
\ \ \ \ 
\bar{\sigma}=\frac{\sqrt{3}en_F}{\sqrt{2}\pi T^2}~~.
\label{scr_ct.eqn}
\end{eqnarray}
Substituting Eq.(\ref{scr_v0.eqn}) into $ F_C $, we obtain,
\begin{equation}
F_C^0 \simeq \sqrt{\frac{2\sqrt{2}\pi e}{3\sqrt{3}}} \frac{N_F^{3/2}}{\sqrt{S}}~.
\label{scr_fc0.eqn}
\end{equation}
Adding the surface energy, we obtain the free energy of the F-fall
within this order approximation,
\begin{eqnarray}
F^0&=&E_s^0+F_C^0 \nonumber \\
&=& \Sigma S+ \left( \frac{4\sqrt{\pi}}{3}
+\frac{2\sqrt{2}\pi e}{3\sqrt{3}} \right) \frac{N_F^{3/2}}{\sqrt{S}}~.
\label{scr_f0.eqn}
\end{eqnarray}
Since the coulomb energy only change the coefficient of the second term,
we can not know the stability of the F-ball within this order
approximation, too.  We therefore need to calculate the next-to-leading
order approximation.

We expand $ V_e $ and $ F_C $ with respect to $ \delta_C/R_1 $ and $
\delta_C/R_2 $ \footnote{Here, $ \delta_C $ is a typical screening
length near the surface of the F-ball, $ \delta_C \simeq \lambda_T c_T
$.},
\begin{eqnarray}
V_e&=&V_e^0+V_e^1+\cdots \nonumber \\
F_C&=&F_C^0+F_C^1+F_C^2+\cdots~.
\label{scr_expand.eqn}
\end{eqnarray}
Since $ F_C^1 $ vanishes, we estimate $ F_C^2 $, expanding $ V_e $ up to
$ V_e^1 $. From Eq.(\ref{scr_eqM.eqn}), we obtain
\begin{equation}
\frac{1}{e^2} \ddot{V_e^1}-
\left( \frac{T^2}{3}+\frac{(V_e^0)^2}{\pi^2} \right) 
V_e^1=-\frac{1}{e^2} \left( \frac{1}{R_1}+\frac{1}{R_2} \right) \dot{V_e^0}~,
\label{scr_eqM1.eqn}
\end{equation}
with the boundary condition,
\begin{equation}
V_e^1 \rightarrow 0 \hspace{1cm} (w \rightarrow \pm \infty)~.
\label{scr_bc1.eqn}
\end{equation}
The solution to Eq.(\ref{scr_eqM1.eqn}) is,
\begin{equation}
V_e^1(w)= \frac{1}{2} \left( \frac{1}{R_1}+\frac{1}{R_2} \right) 
f_2(|w|/\lambda_T)~.
\label{scr_v1.eqn}
\end{equation}
Here, $ f_2(|w|/\lambda_T) $ is defined as,
\begin{equation}
f_2(|w|/\lambda_T)
= 
-\frac{w}{\left| w \right|} \frac{\sqrt{6}\pi}{e} 
\frac{\cosh{(|w|/\lambda_T+c_T)}}{\mbox{sinh}^2(|w|/\lambda_T+c_T)} 
\left\{ f_3(|w|/\lambda_T+c_T)-f_3(c_T) \right\}~,
\label{scr_v1_2.eqn}
\end{equation}
where $ f_3(x) $ is the following function:
\begin{equation}
f_3(x)=\frac{1}{3}\sinh{x}\cosh{x}+\frac{2}{3}\tanh{x}-\frac{1}{3}\mbox{sinh}^2x-x~.
\label{scr_v1_3.eqn}
\end{equation}
Substituting $ V_e^0 $ and $ V_e^1 $ into $ F_C $, we obtain,
\begin{equation}
F_C^2=C_C^1 \int \mbox{d}S~\frac{1}{ R_1R_2 } 
+C_C^2 \int \mbox{d}S~ \left(\frac{1}{R_1}-\frac{1}{R_2}\right)^2~,
\label{scr_fc2.eqn}
\end{equation}
with,
\begin{eqnarray}
C_C^1&=&-\frac{1}{e^2} \int \mbox{d}w~ 
\left\{ w^2 (\dot{V_e^0})^2-\dot{V_e^0} f_2 \right\}
 \simeq -2\pi^{4/3} \left(\frac{2}{3e^2} \right)^{5/4} \lambda^{1/6}v < 0
\nonumber \\
C_C^2&=& \frac{1}{4e^2} \int \mbox{d}w~ \dot{V_e^0} f_2
 \simeq \pi^{4/3} \left(\frac{2}{3e^2} \right)^{5/4}\lambda^{1/6}v~ > 0.
\label{scr_cc.eqn}
\end{eqnarray}
Adding the surface energy, we finally obtain,
\begin{eqnarray}
F^2&=&E_s^2+F_C^2 \nonumber \\
&=&
C^1 \int \mbox{d}S~\frac{1}{ R_1R_2 } 
+C^2 \int \mbox{d}S~ \left(\frac{1}{R_1}-\frac{1}{R_2}\right)^2~,
\label{scr_f2.eqn}
\end{eqnarray}
where $ C^i $ $(i=1,2)$ is the sum of $ C_N^i $ and $ C_C^i $.  Since the
first term is always negative and depends only on the number of the
F-ball, a fragmented state has smaller energy than an F-ball state.
However, the second term is positive in range $ C^2 >0 $, where is,
\begin{eqnarray}
\lambda
\gsim
 e^{\frac{15}{4}}.
\label{lambdapara}
\end{eqnarray}
In this range, an F-ball is stable against the deformation from
spherical shape.  It, however, will fragment due to the tunneling effect
or the thermal fluctuation, since a fragmented state has smaller energy
than an F-ball\footnote{The condition that an F-ball survives until
now will be shown in Ref.\cite{Ara}.}. We therefore conclude that an
electrically charged F-ball is metastable in this range.


\section{Summary and Discussion}

We considered the stability of an F-ball.  We found out that an electrically
neutral F-ball is unstable even if we take into account the curvature
effect.  We notice that this is true even in the absence of the volume
energy.  We then found out that an electrically charged F-ball is
metastable even in thermal bath for the parameter range,
Eq.(\ref{lambdapara}).  

We lastly discuss the allowed region of parameters, $\epsilon, N$, and
$\kappa$.  For the volume energy, which arises from the energy density
difference between the two almost degenerate vacua, $\epsilon$, we have
two constraints from cosmology and from the stability of the
electrically charged F-ball.  From cosmology, the energy density
difference should be large enough to avoid the black hole dominated
universe \cite{Vil}:
\begin{eqnarray}
\epsilon
\gsim
\frac{\Sigma^2}{m_{pl}^2}.
\end{eqnarray} 
 This condition is rewritten as,
\begin{eqnarray}
\epsilon 
\gsim 
\lambda v^4
\left(
\frac{v}{m_{pl}}
\right) ^2.
\label{con1}
\end{eqnarray}
On the other hand, the energy barrier between an F-ball state and a
fragmented state should not vanish due to the volume energy for the
metastability of a charged F-ball.  In order to satisfy this condition,
we have,
\begin{eqnarray}
C^2
\gsim
\epsilon \frac{4 \pi R^3}{3}.
\end{eqnarray}
This condition is rewritten as,
\begin{eqnarray}
\lambda v^4 \lambda^{-\frac{1}{3}} \alpha^{-\frac{5}{4}} N^{-\frac{3}{2}}
\gsim
\epsilon .
\label{con2}
\end{eqnarray}

We next consider a constraint from experiments.  Assuming that F-balls
have sizable contribution to CDM, we obtained a constraint for the
energy of an F-ball in Ref.\cite{Ara2} as,
\begin{eqnarray}
E 
\gsim
10^{25} [GeV]
\ \ \ \ \ 
(\kappa
 \gsim
 10 [GeV]).
\end{eqnarray}
 This condition is rewritten as,
\begin{eqnarray}
N 
\gsim
\left[ \frac{10^{25}}{\kappa [GeV]}\right].
\label{expcon}
\end{eqnarray}

We finally consider how large the symmetry breaking scale, $\kappa$ is?
In order that the conditions, Eq.(\ref{con1}) and Eq.(\ref{con2}), are
compatible, we have,
\begin{eqnarray}
\lambda^{-\frac{1}{3}} \alpha^{-\frac{5}{4}} N^{-\frac{3}{2}}
\gsim
\left(
\frac{v}{m_{pl}}
\right) ^2.
\end{eqnarray}
This condition is rewritten as,
\begin{eqnarray}
N
\lsim
\left(
\frac{m_{pl}}{v}
\right) ^{\frac{4}{3}}
\lambda^{-\frac{2}{9}}
\alpha^{-\frac{5}{6}}.
\label{Nupper}
\end{eqnarray}
In order that the conditions, Eq.(\ref{expcon}) and Eq.(\ref{Nupper}), are
compatible, we have,
\begin{eqnarray}
v 
\lsim
10 \alpha^{-\frac{5}{2}} \lambda^{-\frac{1}{6}} [GeV].
\end{eqnarray}
The symmetry breaking scale seems not to be so far from the electroweak
scale.  These constraints would help us to make a realistic model of an 
F-ball.

\Acknowledgements 
We have benefited from the useful comments by Masahide Yamaguchi. We
would like to thank Kojin Takeda and Hisaki Hatanaka for the helpful
conversations.

\end{document}